\documentclass[aps,prd,amsmath,amssymb,preprintnumbers,nofootinbib,a4paper,11pt]{revtex4}
\pdfoutput=1
\usepackage{amsthm}
\usepackage{graphicx}
	\graphicspath{{./NewFig/}}
\usepackage{url}
\usepackage[bookmarks, pagebackref=false]{hyperref}
\usepackage{color}
	\definecolor{rossoCP3}{cmyk}{0,.88,.77,.40}
		\definecolor{graa}{rgb}{0.8,0.8,0.8}
		\definecolor{blaa}{rgb}{0.2,0.2,0.6}
		\hypersetup{
			colorlinks, 
			bookmarksopen, 
			bookmarksnumbered,
			citecolor=blaa, 		
			linkcolor=rossoCP3,	
			urlcolor=rossoCP3,			
			}
\usepackage{dcolumn}
\usepackage{bm}
\usepackage{bbm}
\usepackage{pxfonts}

\usepackage{epsfig}
\usepackage{placeins}

\usepackage[ margin=5pt, font=small,labelfont=bf, justification=raggedright]{caption}
\usepackage{youngtab}
\usepackage{slashed}
\usepackage{subfig}

\newcommand{\beq}{\begin{eqnarray}}
\newcommand{\eeq}{\end{eqnarray}}

\newcommand{\bmp}{\noindent\begin{minipage}{16cm}}
\newcommand{\emp}{\end{minipage}\vskip 7mm} 

\def\lsim{\mathrel{\rlap{\lower4pt\hbox{\hskip1pt$\sim$}}
    \raise1pt\hbox{$<$}}}                
\def\gsim{\mathrel{\rlap{\lower4pt\hbox{\hskip1pt$\sim$}}
    \raise1pt\hbox{$>$}}}                

\begin{document}

\title{\LARGE \color{rossoCP3} Higher Loop Corrections to the Infrared Evolution of Fermionic Gauge Theories in the RI' Scheme}
 \author{Thomas A. Ryttov}\email{ryttov@cp3.dias.sdu.dk} 
  \affiliation{ {\color{rossoCP3} \rm Center for the Fundamental Laws of Nature} - Jefferson Physical Laboratory \\
  Harvard University - Cambridge, MA 02138 - USA   \\
  {\rm and} \\
{ \color{rossoCP3}  \rm CP}$^{\color{rossoCP3} \bf 3}${\color{rossoCP3}\rm-Origins} \& the Danish Institute for Advanced Study {\color{rossoCP3} \rm DIAS},\\ 
University of Southern Denmark, Campusvej 55, DK-5230 Odense M, Denmark.
}

\begin{abstract}
We study the evolution of the gauge coupling and the anomalous dimension of the mass towards an infrared fixed point for non-supersymmetric gauge theories in the modified regularization invariant, RI', scheme. This is done at the three loop level where all the renormalization group functions have been calculated explicitly. The purpose is to assess the scheme dependence of earlier and similar investigations performed at three and four loop order in the modified minimal subtraction, $\overline{\text{MS}}$, scheme. Our results are of the same order when compared to the $\overline{\text{MS}}$ scheme. 
 \vskip .1cm
{\footnotesize  \it Preprint: CP$^3$-Origins-2013-033  DNRF 90 \& DIAS-2013-33}
 \end{abstract}

\maketitle

\newpage

\section{Introduction} 

Ever since its discovery strong dynamics has continued to pose considerable challenges. Of specific interest is the study of its behavior from high scales to low scales and the type of dynamics that it exhibits in the deep infrared (IR). A set of tools typically used to study such behavior is the renormalization group equations and their associated prediction of the evolution of the gauge coupling. 

An early pioneering step in this direction was first taken by Caswell \cite{Caswell:1974gg} and subsequently by Banks and Zaks \cite{Banks:1981nn} who noted the appearance of an IR fixed point in a certain region of theory space. The fixed point appears just below where asymptotic freedom is lost as one decreases the number of flavors. Lowering the number of flavors even further the quest is now to predict exactly at what critical value the fixed point is lost and where one presumably will enter a chirally broken phase. The region in theory space where one develops an IR fixed point is known as the conformal window. 

Since then there has been a vast amount of work done using truncated Dyson-Schwinger equations to predict the value of the coupling constant that triggers the formation of the chiral condensate \cite{Holdom:1984sk,Yamawaki:1985zg,Appelquist:1986an,Appelquist:1986tr,Appelquist:1987fc,Appelquist:1988yc,Appelquist:1997dc,Brodsky:2008be} while the question of conformality has been studied using the beta function of the theory \cite{Sannino:2004qp,Dietrich:2006cm}. 

Many of the difficulties encountered in the non-supersymmetric case are not present within their $\mathcal{N}=1$ supersymmetric extensions where the conformal window was predicted by Seiberg \cite{Seiberg:1994pq} and later generalized to the case of higher dimensional representations in \cite{Ryttov:2007sr}. These results rely heavily on the existence of the Novikov-Shifman-Vainshtein-Zakharov beta function \cite{Novikov:1983uc,Shifman:1986zi}. Inspired by this a similar all-orders beta function was conjectured for non-supersymmetric theories and used to predict the conformal window \cite{Ryttov:2007cx,Pica:2010mt}.

For non-supersymmetric theories the initial studies were all done in the simplest setup utilizing the two loop beta function. Therefore a more recent approach has been to extend the original analysis to higher orders in perturbation theory. In the non-supersymmetric case the beta function and the anomalous dimension of the mass are known to four loop order in the $\overline{\text{MS}}$ scheme \cite{vanRitbergen:1997va,Vermaseren:1997fq} enabling a study of the stability of previous investigations \cite{Ryttov:2010iz,Pica:2010xq}. Additional work in this direction can be found in \cite{Shrock:2013ca,Shrock:2013pya}. The same question has also been addressed to three loop order in the $\overline{\text{DR}}$ scheme for supersymmetric theories where comparison to exact results can be made \cite{Ryttov:2012qu}. It should be stressed that the higher loop calculations tend to yield a smaller than expected value for the anomalous dimension of the mass at the fixed point. This seems to agree with the majority of the lattice simulations performed in this direction. For a recent review of the latest results see \cite{Lattice}.

It is clear that once the perturbative expansion of the beta function is truncated the question of scheme dependence is inevitable. Studies to address this issue were initiated in \cite{Ryttov:2012ur,Ryttov:2012nt,Shrock:2013uaa} where artificial and well behaved scheme transformations were constructed and used to analyze the stability of the four loop results. However no calculations  in a different and \emph{explicit} scheme has been carried out such that direct comparison with the four loop results in the $\overline{\text{MS}}$ scheme could be made. 

It is the purpose of this paper to undertake such an investigation by studying the evolution of the gauge coupling and the anomalous dimension towards an IR fixed point in the scheme known as the modified regularization invariant, RI', scheme \cite{Martinelli:1994ty}. Here all the renormalization group functions have been calculated to three loop order \cite{Gracey:2003yr}.

In general the beta function and the other renormalization group equations depend on the gauge parameter so it should be stressed that the 't Hooft two loop beta function is universal only within a certain set of schemes \cite{Gross:1973ju,hooft}. Such schemes include the minimal subtraction, MS, scheme \cite{'tHooft:1973mm} and the modified minimal subtraction, $\overline{\text{MS}}$, scheme \cite{Bardeen:1978yd}. Only the one loop beta function is truly universal preserving the property of asymptotic freedom.

In the RI' scheme several of the renormalization group functions depend on the gauge parameter. Hence in order to study the evolution of the gauge coupling and the anomalous dimension of the mass towards the fixed point we must include and make sure that also the gauge parameter is evolving towards the fixed point. For this we should (and will) set up a general framework for tackling such problems. 

Much of the above work has been generalized to multiple fermion representations \cite{Ryttov:2009yw,Ryttov:2010hs} and exceptional gauge groups and spinorial representations \cite{Mojaza:2012zd} while yet a new strategy has been to bound the conformal window using the a-theorem \cite{Antipin:2013qya}. We also mention that the quest for near-conformal dynamics has its roots in Technicolor model building and beyond Standard Model physics, (for a recent review see \cite{Sannino:2009za}).

In Section \ref{notation} we introduce our notation and the various renormalization group functions that is needed while in Section \ref{scheme} we discuss specific schemes, including the RI' scheme, and scheme transformations. We then set up our general method for analyzing the IR fixed points in Section \ref{fixedpoints} and use it explicitly in the RI' scheme in Section \ref{fixedpointsRI'}. Finally we conclude in Section \ref{conclusion}. The two Appendices \ref{App:A}-\ref{App:B} provide all necessary information needed to do the analysis.

\section{Notation}\label{notation}

We consider a gauge theory with gauge group $G$ together with a set of fermions transforming according to an arbitrary representation of the gauge group.  We denote the generators in the representation $r$ of $G$ by $T_r^a,\, a=1\ldots d(G)$. Here $d(r)$ is the
dimension of the representation $r$ and the adjoint
representation is denoted by $G$. The generators are normalized according to
$\text{Tr}\left[T_r^aT_r^b \right] =  T(r) \delta^{ab}$ while the
quadratic Casimir $C_2(r)$ is given by $T_r^aT_r^a =  C_2(r)I$. The
trace normalization factor $T(r)$ and the quadratic Casimir are
connected via $C_2(r) d(r) = T(r) d(G)$. 

The Lagrangian of the theory in the linear covariant gauge is simply written as
\begin{eqnarray}
\mathcal{L} &=& - \frac{1}{4} G_{\mu\nu}^{a}G^{a \mu\nu} + i \bar{\psi}_{f} \slashed{D} \psi^{f} - \frac{1}{2\xi} \left( \partial^{\mu} A_{\mu}^{a} \right)^2 - \bar{c}^a \partial^{\mu} D_{\mu} c^a
\end{eqnarray}
with
\begin{eqnarray}
G_{\mu\nu}^{a} &=& \partial_{\mu} A_{\nu}^{a} - \partial_{\nu} A_{\mu}^{a} - g f^{abc} A_{\mu}^{b} A_{\nu}^{c} \ , \\
D_{\mu}\psi  &=& \partial_{\mu} \psi + i g A_{\mu}^{a} T_{r}^{a} \psi \ , \\
D_{\mu} c^{a} &=& \partial_{\mu} c^{a} - g f^{abc} A_{\mu}^{b} c^{c} \ .
\end{eqnarray}
Here $A_{\mu}^{a}$ is the gauge field, $\psi^{f}$ is the fermion field, $c^{a}$ is the ghost field and $a,b,c=1,\ldots,d(G)$ and $f=1,\ldots,N_f$. Also $\alpha = \frac{g^2}{4\pi}$ is the coupling constant and $\xi$ is the gauge parameter. The above Lagrangian is general and describes the dynamics of an arbitrary gauge theory with $N_f$ sets of Dirac fermions transforming according to an arbitrary representation of the gauge group. In the following when discussing renormalization of the theory we shall stick to this general approach. 

Let us consider the fields and parameters of the above Lagrangian as bare quantities and introduce the renormalized ones according to
\begin{eqnarray}
\left(A^{a}_{\mu}\right)_{\text{bare}} &=&  \sqrt{Z_A} A^{a}_{\mu} \\
\psi_{\text{bare}} &=&  \sqrt{Z_{\psi}} \psi \\
c^{a}_{\text{bare}} &=& \sqrt{Z_{c}} c^{a} \\
g_{\text{bare}} &=& \mu^{\epsilon} Z_g g \\
\xi_{\text{bare}} &=& Z^{-1}_{\xi} Z_{A} \xi
\end{eqnarray}
The scale $\mu$ is introduced to keep the coupling constant $g$ dimensionless in $d=4-2\epsilon$ dimensions. Also $\epsilon$ is the regularizing parameter appearing in dimensional regularization which is to be understood as our method of isolating the various divergencies. Note that in general there are five renormalization constants $Z_{A}, Z_{\psi}, Z_{c}, Z_{g}, Z_{\xi}$. Let us therefore define the following renormalization group functions
\begin{eqnarray}
\gamma_A(\alpha,\xi) = \frac{\partial \ln Z_A}{\partial \ln \mu} \ , \quad \gamma_{\psi}(\alpha,\xi) = \frac{\partial \ln Z_{\psi}}{\partial \ln \mu} \ , \quad \gamma_c(\alpha,\xi) = \frac{\partial \ln Z_c}{\partial \ln \mu}
\end{eqnarray}
\begin{eqnarray}
\beta_{\alpha}(\alpha,\xi) = \frac{\partial \alpha}{\partial \ln \mu} \ , \quad \gamma_{\xi}(\alpha,\xi) = \frac{\partial \ln \xi}{\partial \ln \mu}
\end{eqnarray}
where $\gamma_{A,\psi,c}$ is the anomalous dimension of the gauge field, fermion field and ghost field respectively. Also $\beta_{\alpha}$ is the beta function of the gauge coupling and $\gamma_{\xi}$ is the anomalous dimension of the gauge parameter. One should note that we have made it explicit that in general all of these renormalization group functions depend on both the gauge coupling and the covariant gauge parameter. 

Finally it is simple to check that the anomalous dimension of the gauge field and gauge parameter can be written as functions of the various renormalization group functions according to
\begin{eqnarray}
\gamma_{A} &=& \beta_{\alpha} \frac{\partial \ln Z_A}{\partial \alpha} + \xi \gamma_{\xi} \frac{\partial \ln Z_A}{\partial \xi} \\
\gamma_{\xi} &=& \frac{\beta_{\alpha} \frac{\partial \ln Z_{\xi}}{\partial \alpha} - \gamma_A}{1 - \xi \frac{\partial \ln Z_{\xi}}{\partial \xi}} \label{relation}
\end{eqnarray}

Below we shall be interested in yet another renormalization group function. It is the anomalous dimension of the bilinear operator $\bar{\psi}\psi$. To this end let us define the renormalization constant of the bilinear operator via
\begin{eqnarray}
\left( \bar{\psi}\psi \right)_{bare} &=& Z_{\bar{\psi}\psi} \bar{\psi} \psi
\end{eqnarray}
The associated anomalous dimension of the composite $\bar{\psi}\psi$ operator is then 
\begin{eqnarray}
\gamma(\alpha,\xi) &=& - \frac{\partial \ln Z_{\bar{\psi}\psi}}{\partial \ln \mu}
\end{eqnarray}

\section{Choice of Scheme}\label{scheme}

It is clear that the various renormalization group functions depend on the choice of scheme. First we choose dimensional regularization in $d=4-2\epsilon$ dimensions such that the divergences in the various greens functions appear as poles in $\epsilon$. Second we choose a subtraction procedure. 

The simplest of such subtraction procedures is the one that occurs in the minimal subtraction, MS, scheme \cite{'tHooft:1973mm} for which only the infinity with respect to the regularization is removed. Another more convenient scheme is the modified minimal subtraction, $\overline{\text{MS}}$, scheme \cite{Bardeen:1978yd} where not only the infinite part is subtracted but also a finite constant that includes the Euler-Mascheroni constant. One of the well known and elegant features of the $\overline{\text{MS}}$ scheme is the fact that the beta function of the coupling constant and the anomalous dimension of the mass are both independent of the gauge parameter. 

A third subtraction procedure is a modified version of the regularization invariant, RI, scheme called the RI' scheme \cite{Martinelli:1994ty}. Within this scheme the anomalous dimensions and beta functions have been computed to various orders and for various theories \cite{Franco:1998bm,Chetyrkin:1999pq} with the complete three loop results for any fermionic gauge theory in an arbitrary gauge appearing in \cite{Gracey:2003yr}.

A few words regarding the RI' scheme are in order. Following \cite{Gracey:2003yr} we let $\Sigma_{\psi}(p)$ and $\Sigma_{c}(p)$ denote the bare two-point functions of the fermion and ghost field respectively while
\begin{eqnarray}
\Pi_{\mu\nu}(p) &= & \frac{\Pi_T(p)}{p^2} \left[ \eta_{\mu\nu} - \frac{p_{\mu}p_{\nu}}{p^2} \right] + \Pi_{L}(p) \frac{p_{\mu}p_{\nu}}{(p^2)^2}
\end{eqnarray}
denotes the gluon polarization with $\Pi_T$ and $\Pi_L$ being its transverse and longitudinal parts. The renormalization constants of the fermion, ghost and gluon fields together with the gauge parameter are then defined by the following conditions
\begin{eqnarray}
\lim_{\epsilon \rightarrow 0} \left[ Z_{\psi} \Sigma_{\psi}(p) \right]\vline_{p^2=\mu^2} &=& \slashed{p} \\
\lim_{\epsilon \rightarrow 0} \left[ Z_{c} \frac{\Sigma_{c}(p)}{p^2} \right]\vline_{p^2=\mu^2} &=& 1 \\
\lim_{\epsilon \rightarrow 0} \left[ Z_{A} \Pi_T(p) \right]\vline_{p^2=\mu^2} &=& 1 \\
\lim_{\epsilon \rightarrow 0} \left[ Z_{\xi} \Pi_L(p) \right]\vline_{p^2=\mu^2} &=& 1
\end{eqnarray}
For the fermion wave function renormalization in the RI' scheme the complete finite term is removed and absorbed into the renormalization constant. This is in contrast with the $\overline{\text{MS}}$ scheme where only a specific constant term together with the pole in $\epsilon$ is removed. 

Transversality of the gluon propagator corresponds to the gauge renormalization constant being unity $Z_{\xi}=1$. This was also demonstrated explicitly at three loops in \cite{Gracey:2003yr}. One should note that using Eq. \ref{relation} then leads to
\begin{eqnarray}
\gamma_{A} &=& -\gamma_{\xi}
\end{eqnarray}
In principle one can proceed and renormalize the fermion-gluon and ghost-gluon vertices and check that they yield the same coupling constant definition. However as noted in \cite{Gracey:2003yr} this actually leads to MOM or $\overline{\text{MOM}}$ class of schemes with a different definition of the coupling constant for every vertex. Therefore the coupling constant in the RI' scheme is renormalized in an $\overline{\text{MS}}$ fashion such that only the pole and a single constant term containing the Euler-Mascheroni constant is removed. This yields a beta function of the gauge coupling which is equal to the one in the $\overline{\text{MS}}$ scheme. 

A final word concerns the renormalization of the composite fermion bilinear operator $\bar{\psi}\psi$. Here the renormalization constant is defined via the condition
\begin{eqnarray}
\lim_{\epsilon \rightarrow 0} \left[ Z_{\bar{\psi}\psi} Z_{\psi} \langle \psi(p) (\bar{\psi}\psi)(0) \bar{\psi}(-p) \right] \vline_{p^2=\mu^2} = 1
\end{eqnarray}
This concludes our discussion of the RI' scheme. The complete three loop results for all the above renormalization group functions can be found in \cite{Gracey:2003yr}. Also in Appendix \ref{App:A} we have provided the specific results that will be used below.

\section{Fixed Points}\label{fixedpoints}

One of the most outstanding problems of strongly interacting theories is to elucidate the possible phases they exhibit in the deep IR. Of specific interest is the conformal window, i.e. the region in the number of colors and number of flavors for which the theory flows to an IR fixed point and becomes conformal. To undertake such an analysis the renormalization group equations are an excellent tool. If the theory is to exhibit conformal invariance all the couplings of the theory must run to a fixed point. From the renormalization group point of view we have two dimensionless parameters - the gauge coupling and the gauge parameter. The running of these two parameters are determined by the associated beta functions 
\begin{eqnarray}
\frac{\partial \alpha}{\partial \ln \mu} = \beta_{\alpha}(\alpha,\xi) \qquad \text{and} \qquad \frac{\partial \xi}{\partial \ln \mu} = \beta_{\xi}(\alpha, \xi) = \xi \gamma_{\xi}(\alpha,\xi) 
\end{eqnarray}
The fixed points of the theory are then found by solving the (generally) two coupled equations
\begin{eqnarray}\label{fixedpoint}
\beta_{\alpha}(\alpha_0,\xi_0) = 0 \ , \qquad \beta_{\xi}(\alpha_0,\xi_0) = 0
\end{eqnarray} 
where $\alpha_0$ and $\xi_0$ denote the values of the gauge coupling and the gauge parameter at the fixed point. It is clear that the values of the coupling constant and the gauge parameter at the fixed point are scheme dependent. It is therefore to natural to ask whether there exists a scheme independent and therefore physical quantity at the fixed point. 

Consider now two different schemes $S$ and $S'$. They each have their definitions of the coupling constant and the gauge parameter $(\alpha,\xi)$ and $(\alpha',\xi')$ respectively. Let us assume that the transformation between the two schemes is well behaved and invertible \cite{Ryttov:2012ur,Ryttov:2012nt}. This implies that the coupling constant and the gauge parameter in one scheme will be a smooth function of the coupling constant and the gauge parameter in the other scheme. If we denote the renormalization constant of the bilinear operator $\bar{\psi}\psi$ as $Z_{\bar{\psi}\psi}$ and $Z'_{\bar{\psi}\psi}$ in the schemes $S$ and $S'$ respectively then it follows that
\begin{eqnarray}
\gamma' (\alpha',\xi') &=& \gamma (\alpha,\xi) + \beta_{\alpha}(\alpha,\xi) \frac{\partial \ln F_{\bar{\psi}\psi}}{\partial \alpha} +  \beta_{\xi}(\alpha,\xi) \frac{\partial \ln F_{\bar{\psi}\psi}}{\partial \xi}  
\end{eqnarray}
where
\begin{eqnarray}
F_{\bar{\psi}\psi} &=& \frac{Z_{\bar{\psi}\psi}}{Z'_{\bar{\psi}\psi}}
\end{eqnarray}
It is then clear that the anomalous dimension of the operator $\bar{\psi}\psi$ evaluated at a fixed point is a scheme independent quantity.

In general there will be multiple solutions to the coupled fixed point equations for which we are specifically interested in the IR stable fixed points. To classify the fixed points we linearize the respective renormalization group equations around each of the zeros
\begin{eqnarray}
\frac{\partial}{\partial \ln \mu} 
\left(
\begin{array}{c}
\alpha - \alpha_0 \\
\xi - \xi_0
\end{array}
\right) &=& M 
\left(
\begin{array}{c}
\alpha - \alpha_0 \\
\xi - \xi_0
\end{array}
\right) + O\left( \left(\alpha - \alpha_0 \right)^2,\left(\xi - \xi_0 \right)^2 \right) 
\end{eqnarray}
where 
\begin{eqnarray}
M &=&
\left. \left(
\begin{array}{cc}
\frac{\partial \beta_{\alpha}}{\partial \alpha} & \frac{\partial \beta_{\alpha}}{\partial \xi} \\
\frac{\partial \beta_{\xi}}{\partial \alpha} & \frac{\partial \beta_{\xi}}{\partial \xi}
\end{array}
\right) \right|_{\alpha = \alpha_0,\ \xi = \xi_0}
\end{eqnarray}
The sign of each eigenvalue and whether it is real or complex then classify the fixed point $(\alpha_0,\xi_0)$.

\section{Fixed Points in the RI' Scheme}\label{fixedpointsRI'}

In this section we will investigate the fixed points of gauge theories with fermonic matter within the RI' scheme and compute the scheme independent anomalous dimension of the mass $\gamma$. The beta function and the anomalous dimension of $\bar{\psi}\psi$ were computed at the four loop level in the $\overline{\text{MS}}$ scheme in \cite{vanRitbergen:1997va,Vermaseren:1997fq}. The same high loop accuracy has not quite been reached in the RI' scheme. However all anomalous dimensions and beta functions have been computed directly to three loop order in \cite{Gracey:2003yr}. 

As mentioned above in the RI' scheme the beta function of the coupling constant coincides at this loop order with the one in the $\overline{\text{MS}}$ scheme \cite{Gracey:2003yr}. This also implies that it is independent of the gauge parameter. On the other hand in the RI' scheme the anomalous dimension of the $\bar{\psi}\psi$ operator is gauge dependent whereas in the $\overline{\text{MS}}$ this is also gauge independent. It is therefore crucial that when investigating the fixed points in the RI' scheme one must take full care that both the coupling constant and the gauge parameter have reached the fixed point when evaluating the anomalous dimension $\gamma_{\bar{\psi}\psi}$. This is the reason for our more general treatment of fixed points above. Some of the simplifications enjoyed in the $\overline{\text{MS}}$ scheme are not present in the RI' scheme. 

Following \cite{Gracey:2003yr} we write the beta functions and the anomalous dimension as
\begin{eqnarray}
\beta_{\alpha}\left(\alpha,\xi \right) &=& -b_{\alpha,1} \left( \frac{\alpha}{4\pi} \right)^2 - b_{\alpha,2} \left( \frac{\alpha}{4\pi} \right)^3 -b_{\alpha,3} \left( \frac{\alpha}{4\pi} \right)^4 + O(\alpha^5) \\
\beta_{\xi} \left( \alpha,\xi \right) &=& \xi \left[- b_{\xi,1} \left( \frac{\alpha}{4\pi} \right) - b_{\xi,2} \left( \frac{\alpha}{4\pi} \right)^2- b_{\xi,3} \left( \frac{\alpha}{4\pi} \right)^3 + O(\alpha^4)  \right] \\
\gamma \left(\alpha,\xi \right) &=& c_{1} \left( \frac{\alpha}{4\pi} \right) + c_{2} \left( \frac{\alpha}{4\pi} \right)^2+ c_{3} \left( \frac{\alpha}{4\pi} \right)^3 + O(\alpha^4)
\end{eqnarray}

All of the coefficients are reported in Appendix \ref{App:A}. Here we also report the group factors in Table \ref{factors} for the representations used throughout this paper. They include the fundamental, adjoint, two-indexed symmetric and two-indexed antisymmetric representations. 

The strategy should now be clear. We first find the fixed points of the two coupled beta functions. In general there are several fixed points where some will be discarded on physical grounds. For the IR fixed points we will then evaluate the anomalous dimension of the mass and compare to previous multi loop results obtained in the $\overline{\text{MS}}$ scheme.

For a given representation the range in the number of flavors we are considering is limited from above by the condition that the theory should be asymptotically free
\begin{eqnarray}
N_f &<& \frac{11}{4}\frac{C_2(G)}{T(r)}
\end{eqnarray}
Also the range in the number flavors is limited from below by requiring the value of the coupling constant to be less than order unity in order for our perturbative calculation to make sense. In any event when the coupling constant reaches the critical value \cite{Holdom:1984sk,Yamawaki:1985zg,Appelquist:1986an,Appelquist:1986tr,Appelquist:1987fc}
\begin{eqnarray}
\alpha &\sim& \frac{\pi}{3C_2(r)}
\end{eqnarray}
the dynamics is expected to trigger the formation of the chiral condensate and break chiral symmetry. Also both perturbative and nonperturbative corrections to this one-gluon exchange approximation have been discussed \cite{Appelquist:1988yc,Appelquist:1997dc,Brodsky:2008be}. 

In the analysis of fixed points we are of course limited by perturbation theory. It is clear that when we truncate the expansion of the beta functions at finite order many possible solutions appear due to the higher powers of the gauge coupling and gauge parameter. On physical grounds we shall only consider positive zeros of the gauge coupling beta function but will allow both positive and negative zeros of the gauge parameter beta function. In fact we shall not limit the range at all in which the gauge parameter at the fixed point can take values. In the following we discuss our results.

At two loops the gauge coupling beta function has a double zero at the origin and one zero, $\alpha_{2\ell}$, away from the origin while the gauge parameter beta function has one zero at the origin, $\xi_{2\ell,1}=0$, and three zeros, $\xi_{2\ell,n},\ n=2,3,4$, away from the origin. In the range of flavors we are considering $\alpha_{2\ell}$ is positive. The fixed points then are
\begin{itemize}
\item The first fixed point $(\alpha_{2\ell},\xi_{2\ell,1})$ is a saddle point since the matrix $M$ has one positive and one negative eigenvalue. It is stable in the $\alpha$ direction. This fixed point is therefore only reached along the trajectory $\xi(\mu)=0$ for all scales $\mu$. 
\item The second fixed point $(\alpha_{2\ell},\xi_{2\ell,2})$ is stable from all directions since the eigenvalues of $M$ are positive.  The value of $\xi_{2\ell,2}$ stays just below $-3$ as we decrease the number of flavors from where asymptotic freedom is lost. However the fixed point only exists in a limited range in the number of flavors (in the specific case of the adjoint representation this zero does not exist at all for an integer value of the number of flavors). 
\item The third fixed point $(\alpha_{2\ell},\xi_{2\ell,3})$ is stable from all directions since the eigenvalues of $M$ are positive. The value of $\xi_{2\ell,3}$ is positive in the entire range of flavors we are considering and increases as the number of flavors approaches the value where asymptotic freedom is lost. 
\item The fourth fixed point $(\alpha_{2\ell},\xi_{2\ell,4})$ is a saddle point since $M$ has one positive and one negative eigenvalue. It also only exists for a number of flavors just below the value where asymptotic freedom is lost. In this range $\xi_{2\ell,4}$ is negative and decreases as the number of flavors approaches the value where asymptotic freedom is lost. 
\end{itemize}
At three loops the picture is almost identical to the two loop case. The gauge coupling beta function has an additional zero which is negative and therefore discarded. The gauge parameter beta function has two additional zeros which are complex in the entire range of flavors we are considering and therefore discarded. The remaining fixed points then follow the same pattern as in the two loop analysis. All of the results are summarized in Tables \ref{fundamentalcouplings}-\ref{antisymmetricgammas} in Appendix \ref{App:B}.

One should note that even though the values of the gauge parameter at the first and second fixed points are $\xi_{2\ell,1}=0$ and $\xi_{2\ell,2}\sim-3$ the values of the associated anomalous dimension are almost identical due to the fact that $\gamma(\alpha,0) = \gamma(\alpha,-3)$ at two loops. This changes only slightly at three loops. 

It should also be noted that both the third and fourth zeros, $\xi_{2\ell,3}$ and $\xi_{2\ell,4}$, diverge to plus infinity and minus infinity respectively as the number of flavors approaches the critical value where asymptotic freedom is lost and where perturbation theory is supposed to be accurate. However the value of $\alpha_{2\ell}$ tends to zero mush faster forcing the value of the anomalous dimension to also approach zero not spoiling the consistency of perturbation theory. The situation is identical at three loops. 

Since there is nothing in our analysis that limits the range in which the gauge parameter can take values at the fixed point it is quite satisfactory that such a consistent picture emerges: at the four different IR fixed points the values of the associated anomalous dimension are in good agreement. On the other hand it should also be mentioned that it is unclear whether all of the solutions will persist in the full theory or they might just be an artifact of the truncation of the perturbative expansion. However all of the above considerations give us confidence in the stability of our results. 

Investigating the explicit results we see that a similar type of pattern is observed in the RI' scheme as compared to the $\overline{\text{MS}}$ scheme. When going from two to three loops the value of the anomalous dimension is lowered. This occurs for all the different possible IR fixed points. When explicit values of the anomalous dimension are compared between the two schemes at same loop order we see good agreement for quiet a large range of flavors just below where asymptotic freedom is lost.

\section{Conclusion}\label{conclusion}

We have studied the evolution of a number of gauge theories from the UV to the IR in a region of theory space where they are believed to develop an IR fixed point. This was done utilizing higher order perturbation theory in the RI' scheme. First we had to address how to estimate the anomalous dimension of the mass at the fixed point within the set of schemes in which it depended on the gauge parameter. We found several solutions with a consistent picture emerging and trustable results were then derived. These were of the same order as similar results obtained in the $\overline{\text{MS}}$ scheme \cite{Ryttov:2010iz,Pica:2010xq}.

\acknowledgments
The author would like to thank F. Sannino and R. Shrock for discussions and careful reading of the manuscript. The CP$^3$-Origins centre is partially funded by the Danish National Research Foundation, grant number DNRF90.

\appendix

\section{Renormalization Group Functions in the RI' Scheme}\label{App:A}

The coefficients of the beta function of the gauge coupling are
\begin{eqnarray}
b_{\alpha,1} &=& \frac{11}{3} C_2(G) - \frac{4}{3} T(r) N_f \\
b_{\alpha,2} &=& \frac{34}{3} C_2(G)^2 - 4 C_2(r)T(r)N_f - \frac{20}{3} C_2(G) T(r) N_f \\
b_{\alpha,3} &=& \frac{1}{54}  \left( 2830 C_2(G)^2 T(r) N_f - 2857 C_2(G)^3+1230 C_2(G) C_2(r) T(r) N_f - 316 C_2(r) T(r)^2 N_f^2 \right. \nonumber \\
&& \left. -108 C_2(r)^2 T(r) N_f - 264 C_2(r) T(r)^2 N_f^2 \right)
\end{eqnarray}

Following \cite{Gracey:2003yr} we note that transversality of the gluon propagator corresponds to $Z_{\xi}=1$ and therefore $\gamma_{\xi} = - \gamma_{A}$. This was checked explicitly at the three loop level in the RI' scheme. Therefore we can write the beta function of the gauge parameter as $\beta_{\xi} = \xi \gamma_{\xi} = -\xi \gamma_{A}$. The coefficients of the beta function of the gauge parameter then are, following \cite{Gracey:2003yr}
\begin{eqnarray}
b_{\xi,1} &=& \frac{1}{6} \left( 8 T(r) N_f - \left(13-3\xi \right)C_2(G) \right) \\
b_{\xi,2} &=& -\frac{1}{216} \left[ \left( 27\xi^3 - 90\xi^2 - 426 \xi +3727 \right)C_2(G)^2 + \left( 72\xi^2 + 240 \xi - 3616 \right)C_2(G)T(r)N_f \right. \nonumber \\
&& \left. -864C_2(r) T(r) N_f + 640 T(r)^2 N_f^2 \right] \\
b_{\xi,3} &=& \frac{1}{7776} \left[ 51200 T(r)^3 N_f^3 - 15552 C_2(r)^2 T(r) N_f + \left( 331776\zeta(3) - 487296 \right)C_2(r) T(r)^2 N_f^2 \right. \nonumber \\
&&  - \left( 486 \xi^5  + 3078\xi^4 + 10260 \xi^3 - 1458\zeta(3)\xi^2 - 25965\xi^2 + 86184\zeta(3) \xi - 173406\xi \right. \nonumber \\
&&  - 175446\zeta(3)  + 2127823 \big)C_2(G)^3 - \left( 648\xi^4 + 216 \xi^3 + 47808\xi^2 + 10368\zeta(3)\xi +126480\xi \right. \nonumber\\
&& - 254016\zeta(3)  - 2501184 \big) C_2(G)^2 T(r) N_f - \left(7776\xi^2 - 62208\zeta(3) \xi +71280\xi + 725760\zeta(3) \right. \nonumber \\
&& - 1131408 \big) C_2(G) C_2(r) T(r) N_f + \left( 11520 \xi^2 +19200\xi - 165888\zeta(3) \right. \nonumber \\
&& \left. - 751680 \big)C_2(G) T(r)^2 N_f^2 \right]
\end{eqnarray}

The coefficients of the anomalous dimension of the mass are
\begin{eqnarray}
c_1 &=& 6 C_2(r) \\
c_2 &=& \frac{1}{3}\left[ \left( 185 + 9 \xi + 3 \xi^2 \right) C_2(G) + 9 C_2(r) - 52T(r) N_f \right]C_2(r) \\
c_3 &=& - \frac{1}{108} \left[ \left( 108\xi^3 + 324 \xi^2 - 1944 - 19008\zeta(3) \right) C_2(G) C_2(r) - \left( 117428 + 5634\xi +1905\xi^2 + 405 \xi^3  \right. \right. \nonumber \\
&& + 54\xi^4 - 28512\zeta(3) \big) C_2(G)^2 + \left( 480 \xi^2 + 2088\xi + 62960 \right) C_2(G) T(r) N_f - 13932 C_2(r)^2 \nonumber \\
&&+  \left( 16632- 3456\zeta(3) \right)  C_2(r) T(r) N_f - 6848T(r)^2 N_f^2 \big] C_2(r)
\end{eqnarray}

\begin{table}[h]
\begin{center}
    \begin{tabular}{c||ccc }
    r & $ \quad T(r) $ & $\quad C_2(r) $ & $\quad
d(r) $  \\
    \hline \hline
    $ \tiny\yng(1) $ & $\quad \frac{1}{2}$ & $\quad\frac{N^2-1}{2N}$ &\quad
     $N$  \\
        $\text{$G$}$ &\quad $N$ &\quad $N$ &\quad
$N^2-1$  \\
        $\tiny\yng(2)$ & $\quad\frac{N+2}{2}$ &
$\quad\frac{(N-1)(N+2)}{N}$
    &\quad$\frac{N(N+1)}{2}$    \\
        $\tiny\yng(1,1)$ & $\quad\frac{N-2}{2}$ &
    $\quad\frac{(N+1)(N-2)}{N}$ & $\quad\frac{N(N-1)}{2}$
    \end{tabular}
    \end{center}
\caption{Relevant group factors for the representations used
throughout this paper.}\label{factors}
    \end{table}

\clearpage

\section{Tables}\label{App:B}

\begin{table}[h]
\caption{\footnotesize{Values of the infrared zeros in $\alpha$ and $\xi$ of the SU($N$) beta functions with $N_f$ fermions in the fundamental
representation, for $N=2,3,4$. They are calculated at $n$-loop order, and denoted as
$\alpha_{n\ell}$ and $\xi_{n\ell}$.}}
\begin{center}
\begin{tabular}{|c|c||c|c|c|c|c||c|c|c|c|c|} \hline\hline
$N$ & $N_f$ & $\alpha_{2\ell}$ & $\xi_{2\ell,1}$ & $\xi_{2\ell,2}$ & $\xi_{2\ell,3}$ & $\xi_{2\ell,4}$ & $\alpha_{3\ell}$ 
& $\xi_{3\ell,1}$ & $\xi_{3\ell,2}$ & $\xi_{3\ell,3}$ & $\xi_{3\ell,4}$  \\ \hline
 2  &  7  &  2.83    & 0    & -          & 3.97  & -       & 1.05   & 0   & -         & 2.63 & -         \\
 2  &  8  &  1.26    & 0    & -          & 4.92  & -        & 0.688 & 0  & -         & 3.23 & -         \\
 2  &  9  &  0.595  & 0    & -          & 6.50  & -        & 0.418 & 0  & -3.81 & 4.32 & -6.50 \\
 2  & 10 &  0.231  & 0    & -3.24  & 10.1 & -10.2 & 0.196 & 0  & -3.15 & 6.80 & -9.42 \\
 \hline
 3  & 10  &  2.21      & 0    &  -         &  3.92   & -         & 0.764   &  0 & -          & 2.57 & -          \\
 3  & 11  &  1.23      & 0   &  -         &   4.46   & -         & 0.579   &  0 & -          & 2.90 & -          \\
 3  & 12  &  0.754    & 0   &  -          & 5.15   & -          & 0.435   &  0 & -         & 3.35 &  -          \\
 3  & 13  &  0.468    & 0   &  -          &  6.12  & -          & 0.317   &  0 & -4.22 & 4.02 &  -5.86  \\
 3  & 14  &  0.278    & 0   &  -3.67   &  7.64  & -6.86 & 0.215   &  0 & -3.42 & 5.08 &  -7.51  \\
 3  & 15  &  0.143    & 0   &  -3.22   &  10.4  & -10.6 & 0.123   &  0 & -3.14 & 7.05 & -9.66    \\
 3  & 16  &  0.0416  & 0   &  -3.04  &  19.6  &  -20.3 & 0.0397 &  0 & -3.03 &13.29 & -15.9  \\ 
\hline
 4  & 13  &  1.85     & 0    & -           &  3.87    &  -          & 0.604   & 0  & -         & 2.53 & -         \\
 4  & 14  &  1.16     & 0    & -           &  4.25    &  -          & 0.489   & 0  & -         & 2.75 & -         \\
 4  & 15  &  0.783   & 0    & -           &  4.69    &  -          & 0.397   & 0  & -         & 3.02 & -         \\
 4  & 16  &  0.546   & 0    & -           &  5.23     & -          & 0.320   & 0  & -         & 3.39 & -         \\
 4  & 17  &  0.384   & 0    & -            &  5.93    & -          & 0.254   & 0  & -4.69 & 3.87 & -5.30  \\
 4  & 18  &  0.266   & 0    & -4.38    &  6.88    & -5.17  & 0.194   & 0  & -3.68 & 4.54 & -6.80  \\
 4  & 19  &  0.175   & 0    & -3.49    &  8.28    & -7.80  & 0.140   & 0  & -3.31 & 5.51 & -8.02  \\
 4  & 20  &  0.105    & 0   & -3.21    &  10.6    & -10.7  & 0.0907 & 0  & -3.13 & 7.13 & -9.74  \\
 4  & 21  &  0.0472  & 0   & -3.07    &  15.8    & -16.4  & 0.0441 & 0  & -3.05 & 10.7 & -13.4 \\ 
\hline\hline
\end{tabular}
\end{center}
\label{fundamentalcouplings}
\end{table}

\begin{table}
\caption{\footnotesize{Values of the anomalous dimension of the $\bar{\psi}\psi$ operator with $N_f$ fermions in the fundamental
representation, for $N=2,3,4$. They are calculated at $n$-loop order, and denoted as $\gamma_{n\ell}$. We also include the values in the $\overline{\text{MS}}$ scheme.}}
\begin{center}
\begin{tabular}{|c|c||c|c|c|c||c|c|c|c||c|c|c|} \hline\hline
&  &   \multicolumn{4}{c||}{RI'} & \multicolumn{4}{c||}{RI'} & \multicolumn{3}{c|}{$\overline{\text{MS}}$}  \\
\hline
$N$ & $N_f$ & $\gamma_{2\ell,1}$ & $\gamma_{2\ell,2}$ & $\gamma_{2\ell,3}$ & $\gamma_{3\ell,4}$ & $\gamma_{3\ell,1}$ & $\gamma_{3\ell,2}$ & $\gamma_{3\ell,3}$ & $\gamma_{3\ell,4}$ & $\gamma_{2\ell}$ & $\gamma_{3\ell}$ & $\gamma_{4\ell}$   \\ \hline
 2  &  7  &  3.49      & -             & 5.60  &  -           & 0.671 & -            & 1.17 & -               & 2.67     & 0.457   & 0.0325 \\
 2  &  8  &  0.872    & -             & 1.46  &  -           & 0.312 & -            & 0.546 & -            & 0.752   & 0.272   & 0.204   \\
 2  &  9  &  0.293    & -             & 0.501 & -           & 0.166 & 0.169   & 0.285 & 0.224   &0.275    & 0.161   & 0.157   \\
 2  & 10 &  0.0924  & 0.0928 & 0.159 & 0.129  & 0.0740 & 0.0741 & 0.126 & 0.114 &0.0910 & 0.0738 & 0.0748 \\
 \hline
 3  & 10  &   5.62     &  -             &  8.97      &   -             & 1.04    & -             & 1.76      & -          & 4.19           & 0.647   & 0.156  \\
 3  & 11  &   1.99     &  -             &  3.27      &   -             & 0.571  & -            & 0.989    &  -          & 1.61          & 0.439   & 0.250   \\
 3  & 12  &   0.888   &  -             &  1.49      &   -             & 0.354  & -            & 0.613    &  -          & 0.773       & 0.312   &  0.253  \\
 3  & 13  &   0.439   &  -             &   0.749   &   -            & 0.232   & 0.242   & 0.398    & 0.292  & 0.404       & 0.220   &   0.210 \\
 3  & 14  &   0.221    & 0.226    &   0.380   &   0.273   & 0.148   & 0.149   & 0.253    & 0.217  & 0.212       & 0.146   &  0.147  \\
 3  & 15  &   0.101    & 0.101    &   0.174   &   0.143   & 0.0828 & 0.0828 & 0.140    & 0.128  & 0.0.0997 & 0.0826 &  0.0836 \\
 3  & 16  &   0.0272  & 0.0272 &    0.0466 &  0.0426 & 0.0258 & 0.0258 & 0.0436 & 0.0417& 0.0272    & 0.0258  & 0.0259  \\ 
\hline
 4  & 13  & 7.33     &  -              & 11.7       &  -           & 1.27     & -             & 2.12    & -             & 5.38       & 0.755    & 0.192  \\
 4  & 14  & 3.13     &  -              &  5.09      &  -           & 0.784   & -             & 1.34    & -             & 2.45       & 0.552    & 0.259 \\
 4  & 15  & 1.59     &  -              &  2.64      &  -           & 0.523   & -             & 0.900  & -             & 1.32       & 0.420    & 0.281 \\
 4  & 16  & 0.892   &  -              &  1.50      &  -           & 0.368   & -             & 0.634  & -             & 0.778    & 0.325    & 0.269 \\
 4  & 17  & 0.528   &  -              &  0.898    & -            & 0.267   & 0.292    & 0.459  &  0.314   & 0.481    & 0.251    & 0.234 \\
 4  & 18  & 0.318   &  0.339     &  0.546    & 0.356   & 0.194   & 0.196    & 0.331  &  0.270   & 0.301    & 0.189    &  0.187 \\
 4  & 19  & 0.189   &  0.192     &  0.326    & 0.244   & 0.136   & 0.136    & 0.230  &  0.202    & 0.183    & 0.134   & 0.136 \\
 4  & 20  & 0.104   &  0.104     &  0.179    & 0.147   & 0.0856 & 0.0856  & 0.145  &  0.133    & 0.102   & 0.0854 & 0.0865 \\
 4  & 21  & 0.0441 &  0.0442   &  0.0757 & 0.0675 & 0.0407 & 0.0407 & 0.0688 &  0.0651 & 0.0440 & 0.0407 & 0.0409 \\ 
\hline\hline
\end{tabular}
\end{center}
\label{fundamentalgammas}
\end{table}

\begin{table}
\caption{\footnotesize{Values of the infrared zeros in
$\alpha$ and $\xi$ of the SU($N$) beta functions with $N_f=2$ fermions in the adjoint
representation, for $N=2,3,4$. They are calculated at $n$-loop order, and denoted as
$\alpha_{n\ell}$ and $\xi_{n\ell}$.  }}
\begin{center}
\begin{tabular}{|c|c||c|c|c|c|c||c|c|c|c|c|} \hline\hline
$N$ & $N_f$ & $\alpha_{2\ell}$ & $\xi_{2\ell,1}$ & $\xi_{2\ell,2}$ & $\xi_{2\ell,3}$ & $\xi_{2\ell,4}$ & $\alpha_{3\ell}$ 
& $\xi_{3\ell,1}$  & $\xi_{3\ell,2}$ & $\xi_{3\ell,3}$  & $\xi_{3\ell,4}$   \\ \hline
 2 &  2  &  0.628 & 0 & - & 6.72  & -  & 0.459  & 0 & -4.14 & 4.20  & -5.99  \\
 3 &   2 &  0.419 & 0 & - & 6.72  & -  & 0.306  & 0 & -4.14 & 4.20 &  -5.99   \\
 4 &  2  &  0.314 & 0 & - & 6.72  &  - & 0.229  & 0 & -4.14 & 4.20 &  -5.99  \\
\hline\hline
\end{tabular}
\end{center}
\label{adjointcouplings}
\end{table}

\begin{table}
\caption{\footnotesize{Values of the anomalous dimension of the $\bar{\psi}\psi$ operator with $N_f=2$ fermions in the adjoint
representation, for $N=2,3,4$. They are calculated at $n$-loop order, and denoted as $\gamma_{n\ell}$. We also include the values in the $\overline{\text{MS}}$ scheme.}}
\begin{center}
\begin{tabular}{|c|c||c|c|c|c||c|c|c|c||c|c|c|} \hline\hline
&  &   \multicolumn{4}{c||}{RI'} & \multicolumn{4}{c||}{RI'} & \multicolumn{3}{c|}{$\overline{\text{MS}}$}  \\
\hline
$N$ & $N_f$ & $\gamma_{2\ell,1}$ & $\gamma_{2\ell,2}$ & $\gamma_{2\ell,3}$ & $\gamma_{2\ell,4}$  & $\gamma_{3\ell,1}$ & $\gamma_{3\ell,2}$ & $\gamma_{3\ell,3}$ & $\gamma_{3\ell,4}$ & $\gamma_{2\ell}$ & $\gamma_{3\ell}$ & $\gamma_{4\ell}$  \\ \hline
 2  &  2  &  0.900    & - &  1.55  &  - & 0.593  & 0.616 & 0.956  & 0.758  & 0.820 & 0.543 & 0.500  \\
 3  &  2  &  0.900    & - &  1.55  &  - & 0.593  & 0.616 & 0.956 &  0.758  & 0.820 & 0.543 & 0.523  \\
 4  &  2  &  0.900    & - &  1.55  &  - & 0.593  & 0.616 & 0.956 & 0.758   & 0.820 & 0.543 & 0.532 \\
\hline\hline
\end{tabular}
\end{center}
\label{adjointgammas}
\end{table}

\begin{table}
\caption{\footnotesize{Values of the infrared zeros in
$\alpha$ and $\xi$ of the SU($N$) beta functions with $N_f$ fermions in the two-indexed symmetric
representation, for $N=3,4$. They are calculated at $n$-loop order, and denoted as
$\alpha_{n\ell}$ and $\xi_{n\ell}$. }}
\begin{center}
\begin{tabular}{|c|c||c|c|c|c|c||c|c|c|c|c|} \hline\hline
$N$ & $N_f$ & $\alpha_{2\ell}$ & $\xi_{2\ell,1}$ & $\xi_{2\ell,2}$ & $\xi_{2\ell,3}$ & $\xi_{2\ell,4}$ & $\alpha_{3\ell}$ 
& $\xi_{3\ell,1}$ &  $\xi_{3\ell,2}$ &  $\xi_{3\ell,3}$ &  $\xi_{3\ell,4}$ \\ \hline
 3 &  2  &  0.842   &  0   & -            & 5.45 & -         & 0.500    & 0  &  -        & 3.27  & -          \\
 3 &   3 &  0.0849 &  0  &  -3.12    & 13.7 & -13.9 & 0.0790  & 0 & -3.07  & 9.11  & -11.7  \\
\hline
 4 &  2  &  0.967   &  0   & -           & 4.92  & -         & 0.485 & 0  & -          & 2.92 & -         \\
 4 &  3  &  0.152   &  0   & -3.46   & 9.11  & -8.32 & 0.129 & 0  & -3.29  & 5.86 & -8.30  \\
\hline\hline
\end{tabular}
\end{center}
\label{symmetriccouplings}
\end{table}

\begin{table}
\caption{\footnotesize{Values of the anomalous dimension of the $\bar{\psi}\psi$ operator with $N_f$ fermions in the two-indexed symmetric
representation, for $N=3,4$. They are calculated at $n$-loop order, and denoted as $\gamma_{n\ell}$. We also include the values in the $\overline{\text{MS}}$ scheme.}}
\begin{center}
\begin{tabular}{|c|c||c|c|c|c||c|c|c|c||c|c|c|} \hline\hline
&  &   \multicolumn{4}{c||}{RI'} & \multicolumn{4}{c||}{RI'} & \multicolumn{3}{c|}{$\overline{\text{MS}}$}  \\
\hline
$N$ & $N_f$ & $\gamma_{2\ell,1}$ & $\gamma_{2\ell,2}$  & $\gamma_{2\ell,3}$ & $\gamma_{2\ell,4}$ & $\gamma_{3\ell,1}$ & $\gamma_{3\ell,2}$ & $\gamma_{3\ell,3}$ & $\gamma_{3\ell,4}$ & $\gamma_{2\ell}$ & $\gamma_{3\ell}$ & $\gamma_{4\ell}$  \\ \hline
 3  &  2  &   2.96     &  -           & 5.03    & -          & 1.70    & -           & 2.57  &   -          & 2.44     & 1.28   & 1.12    \\
 3  &  3  &   0.145   &  0.145  & 0.250  & 0.245 & 0.133 & 0.133  & 0.219 & 0.215  & 0.144  & 0.133 & 0.133 \\
 \hline
 4  &  2  &  6.24   & -            & 10.4    & -         & 3.19     & -            & 4.62   & -          & 4.82   &   2.08  & 1.79    \\
 4  &  3  &  0.395 & 0.399   & 0.685 & 0.511 & 0.319  & 0.320   & 0.520 & 0.488 & 0.381 & 0.313 &  0.315 \\
\hline\hline
\end{tabular}
\end{center}
\label{symmetricgammas}
\end{table}

\begin{table}
\caption{\footnotesize{Values of the infrared zeros in
$\alpha$ and $\xi$ of the SU($N$) beta functions with $N_f$ fermions in the two-indexed antisymmetric
representation, for $N=4$. They are calculated at $n$-loop order, and denoted as
$\alpha_{n\ell}$ and $\xi_{n\ell}$.}}
\begin{center}
\begin{tabular}{|c|c||c|c|c|c|c||c|c|c|c|c|} \hline\hline
$N$ & $N_f$ & $\alpha_{2\ell}$ & $\xi_{2\ell,1}$ & $\xi_{2\ell,2}$ & $\xi_{2\ell,3}$ & $\xi_{2\ell,4}$ & $\alpha_{3\ell}$ 
& $\xi_{3\ell,1}$ &  $\xi_{3\ell,2}$ &  $\xi_{3\ell,3}$ &  $\xi_{3\ell,4}$   \\ \hline
 4  &  6   &  2.16    & 0  & -          &  3.91  &  -        & 0.664    &  0  & -          & 2.48  & -        \\
 4  &   7  &  0.890  & 0  & -         &  4.66  &  -         & 0.437    &  0  & -          & 2.94 &  -        \\
 4  &  8   &  0.449  & 0  & -         &  5.71  &  -         & 0.287    &  0  & -          & 3.64 &  -        \\
 4  &  9   &  0.225  & 0  & -3.86 &  7.47  &  -6.28 & 0.174    &  0  & -3.53  & 4.88 &  -7.22 \\
 4 &  10 &  0.090  & 0  & -3.17 &  11.5  &  -11.6 & 0.0804  & 0  & -3.11  & 7.65 &  -10.3 \\
\hline\hline
\end{tabular}
\end{center}
\label{antisymmetriccouplings}
\end{table}

\begin{table}
\caption{\footnotesize{Values of the anomalous dimension of the $\bar{\psi}\psi$ operator with $N_f$ fermions in the two-indexed antisymmetric
representation, for $N=3,4$. They are calculated at $n$-loop order, and denoted as $\gamma_{n\ell}$. We also include the values in the $\overline{\text{MS}}$ scheme.}}
\begin{center}
\begin{tabular}{|c|c||c|c|c|c||c|c|c|c||c|c|c|} \hline\hline
&  &   \multicolumn{4}{c||}{RI'} & \multicolumn{4}{c||}{RI'} & \multicolumn{3}{c|}{$\overline{\text{MS}}$}  \\
\hline
$N$ & $N_f$ & $\gamma_{2\ell,1}$ & $\gamma_{2\ell,2}$  & $\gamma_{2\ell,3}$ & $\gamma_{2\ell,4}$ & $\gamma_{3\ell,1}$ & $\gamma_{3\ell,2}$ & $\gamma_{3\ell,3}$ & $\gamma_{3\ell,4}$ & $\gamma_{2\ell}$ & $\gamma_{3\ell}$ & $\gamma_{4\ell}$  \\ \hline
 4  &  6  &  13.7    &  -           & 21.8    & -            & 2.57   & -          & 4.01    & -           & 9.78    & 1.38   & 0.293 \\
 4  &  7  &   2.73   &  -           &  4.52   & -            & 0.942 & -          & 1.56   &  -           & 2.19    & 0.695 & 0.435 \\
 4  &  8  &   0.904 & -            &  1.54   & -            & 0.449 & -          & 0.756 & -           & 0.802  & 0.402 & 0.368 \\
 4  &  9  &   0.348 &  0.359  &  0.600 & 0.414  & 0.234  & 0.236 & 0.394 & 0.336  & 0.331  & 0.228 & 0.232 \\
 4  & 10  &  0.118 &  0.119  &  0.204 & 0.170  & 0.101 & 0.101  & 0.171 & 0.159  & 0.117 & 0.101  & 0.103 \\
\hline\hline
\end{tabular}
\end{center}
\label{antisymmetricgammas}
\end{table}

\end{document}